\documentclass[a4paper,11pt]{article}
\usepackage{pos}
\usepackage{graphicx}

\usepackage{subcaption} 
\title{Astro-COLIBRI: A Comprehensive Platform for Real-Time Multi-Messenger Astrophysics
}
\ShortTitle{Astro-COLIBRI}

\author*[a]{Fabian Sch\"ussler}
\author[a]{B. Cornejo}
\author[a]{M. Costa}
\author[a]{I. Jaroschewski}
\author[a]{W. Kiendrébéogo}
\onbehalf{on behalf of the Astro-COLIBRI team}
\affiliation[a]{IRFU, CEA, Université Paris-Saclay, Gif-sur-Yvette, France}

\emailAdd{fabian.schussler@cea.fr}
\emailAdd{astro.colibri@gmail.com}

\abstract{The detection of transient phenomena such as Gamma-Ray Bursts (GRBs), Fast Radio Bursts (FRBs), stellar flares, novae, and supernovae—alongside novel cosmic messengers like high-energy neutrinos and gravitational waves—has transformed astrophysics in recent years. Maximizing the discovery potential of multi-messenger and multi-wavelength follow-up observations, as well as serendipitous detections, requires a tool that rapidly compiles and contextualizes relevant information for each new event. We present Astro-COLIBRI, an advanced platform designed to meet this challenge.

Astro-COLIBRI integrates a public RESTful API, real-time databases, a cloud-based alert system, and user-friendly clients (a website and mobile apps for iOS and Android). It processes astronomical alerts from multiple streams in real time, filtering them based on user-defined criteria and placing them in their multi-wavelength and multi-messenger context. The platform offers intuitive data visualization, a quick summary of relevant event properties, and an assessment of observing conditions at numerous observatories worldwide.

We here describe its architecture, data resources, and main functionalities. We highlight the automatic collection of photometric data from a variety of large scale optical surveys, a recently added feature that significantly improves the capabilities of the Astro-COLIBRI platform.}

\ConferenceLogo{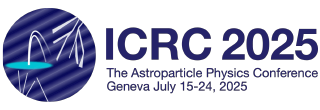}

\FullConference{39th International Cosmic Ray Conference (ICRC2025)\\
 15–24 July 2025\\
Geneva, Switzerland\\}

\begin{document}
\maketitle

\section{Introduction}
\label{sec:intro}
The modern era of time-domain and multi-messenger (TDAMM) astrophysics is characterized by a deluge of data. Current and upcoming observatories across the electromagnetic spectrum, coupled with detectors for gravitational waves, high-energy neutrinos, and cosmic rays, are discovering transient phenomena at an unprecedented rate. These events, from supernova explosions and gamma-ray bursts to the mergers of compact objects and flares of active galactic nuclei, offer unique insights into the most extreme processes in the universe. However, the sheer volume, rapidity, and variety of alerts from dozens of independent distribution streams present a formidable challenge for astronomers seeking to identify, prioritize, and follow up on the most scientifically compelling events.

To address this challenge, we developed Astro-COLIBRI, a comprehensive platform designed to streamline the workflow of transient astronomy for both professional and amateur communities~\citep{Reichherzer2021}. Its core mission is to aggregate transient alerts from all available public channels, filter them according to user-specific criteria, and place them into their broader scientific context in real time. The platform is accessible via a website, dedicated iOS and Android applications, and via a public REST API, ensuring that users can stay informed regardless of their location, workflow, and existing software environment.

\subsection{Astro-COLIBRI System Architecture and Data Flow}
The platform is built on a modern, modular, and cloud-based architecture optimized for performance, reliability, and scalability. The main components are:
\begin{itemize}
    \item \textbf{RESTful API:} The system's backbone is a public RESTful API developed in Python using the Flask framework. It handles all core logic, data processing, and communication with external services. It is running with extremely high availability on a fully cloud-based cluster. 
    \item \textbf{Databases:} Astro-COLIBRI utilizes a dual-database system. A MongoDB instance serves as the main persistent storage for all event and source data. A Firebase Firestore real-time database is used to stream live data directly to the user clients, ensuring the interface is always up-to-date with negligible latency.
    \item \textbf{Real-Time Listeners:} Dedicated servers constantly listen to a multitude of alert streams, including VOEvents from brokers like the General Coordinates Network (GCN), notices from the Transient Name Server (TNS), the Central Bureau for Astronomical Telegrams (CBAT), and other alert distribution channels.
    \item \textbf{Cross-Platform Clients:} To ensure wide accessibility, the user interfaces—website and mobile apps—are built from a single codebase using the Flutter framework. This allows for rapid development and a consistent user experience across all devices.
    \item \textbf{Cloud-Based Alerting:} A flexible notification system using Firebase Cloud Messaging (FCM) sends real-time push alerts to users' mobile devices based on their personalized filter criteria.
\end{itemize}

The data flows from monitoring observatories, through alert brokers, into the Astro-COLIBRI listeners. The API then processes each alert, enriching it with contextual information from astronomical databases (like SIMBAD, NED, and Gaia) before storing it and pushing it out to users in real time.

\subsection{Core Functionalities}

One of the platform's primary functions is to provide a top-level summary of every transient event. Astro-COLIBRI's backend systems continuously monitor and parse incoming communications from sources like GCN, TNS, CBAT, and various survey-specific alert streams. While keeping the sources of information separate, it intelligently associates individual messages from each of these brokers related to a single event, presenting an up-to-date view of its discovery and follow-up history. This saves researchers valuable time that would otherwise be spent manually collating information from disparate sources.

Maximizing the scientific return from a new transient often hinges on the ability to conduct rapid follow-up observations. Astro-COLIBRI provides a powerful suite of visibility tools to facilitate this crucial step. For any event, the platform can instantly calculate and display its visibility from a list containing the most relevant observatories around the world, complemented with a database of over 2600 professional and amateur observatories compiled by the International Astronomical Union (IAU), as well as any user-defined custom location. These calculations, presented as intuitive altitude vs time charts, account for factors such as local time, sun and moon positions, and moon illumination, allowing observers to quickly determine the feasibility and optimal timing for follow-up observations. The detailed observability windows are also available in easy-to-use JSON format and can thus be incorporated in existing observation scheduling software.

A long-term as well as a multi-observatory visibility figure complete this feature. They allow for efficient observation planning, facilitating the choice of the best moment and the best instrument to use for a successful campaign.

Beyond these core features, Astro-COLIBRI enriches each event by providing crucial multi-wavelength and multi-messenger context. It allows to search the event's location in major astronomical databases (e.g., NED, SIMBAD, VizieR) to identify potential host galaxies or known counterparts. A recently added feature allows to calculate a color-color diagram using the WISE infrared survey. Overlaying common classification of galaxy properties allows to rapidly assess the type of host. For high-energy neutrino or gravitational wave events, it provides access to specialized tools, such as optimized tiling strategies for follow-up observations with wide field-of-view instruments using the \texttt{tilepy} cloud-based API~\citep{SeglarArroyo2024}. 

Astro-COLIBRI is available as a web application and smartphone application. The latter enables the use of a highly-configurable push notification system which ensures that users are alerted to new events that match their specific scientific interests the moment they are reported.

While these features provide essential contextual and logistical information, a direct, time-resolved view of an event's brightness is fundamental for its physical characterization. The photometric evolution, or lightcurve, is a key diagnostic tool for classifying a transient and understanding the physics driving its emission. To provide users with this vital information, we have developed and integrated an automated lightcurve generation feature. In the following we detail the implementation of this new feature, from the data sources used to the final products delivered to the user.

\begin{figure}
\centering
\begin{subfigure}{.65\textwidth}
  \centering
  \includegraphics[width=.95\linewidth]{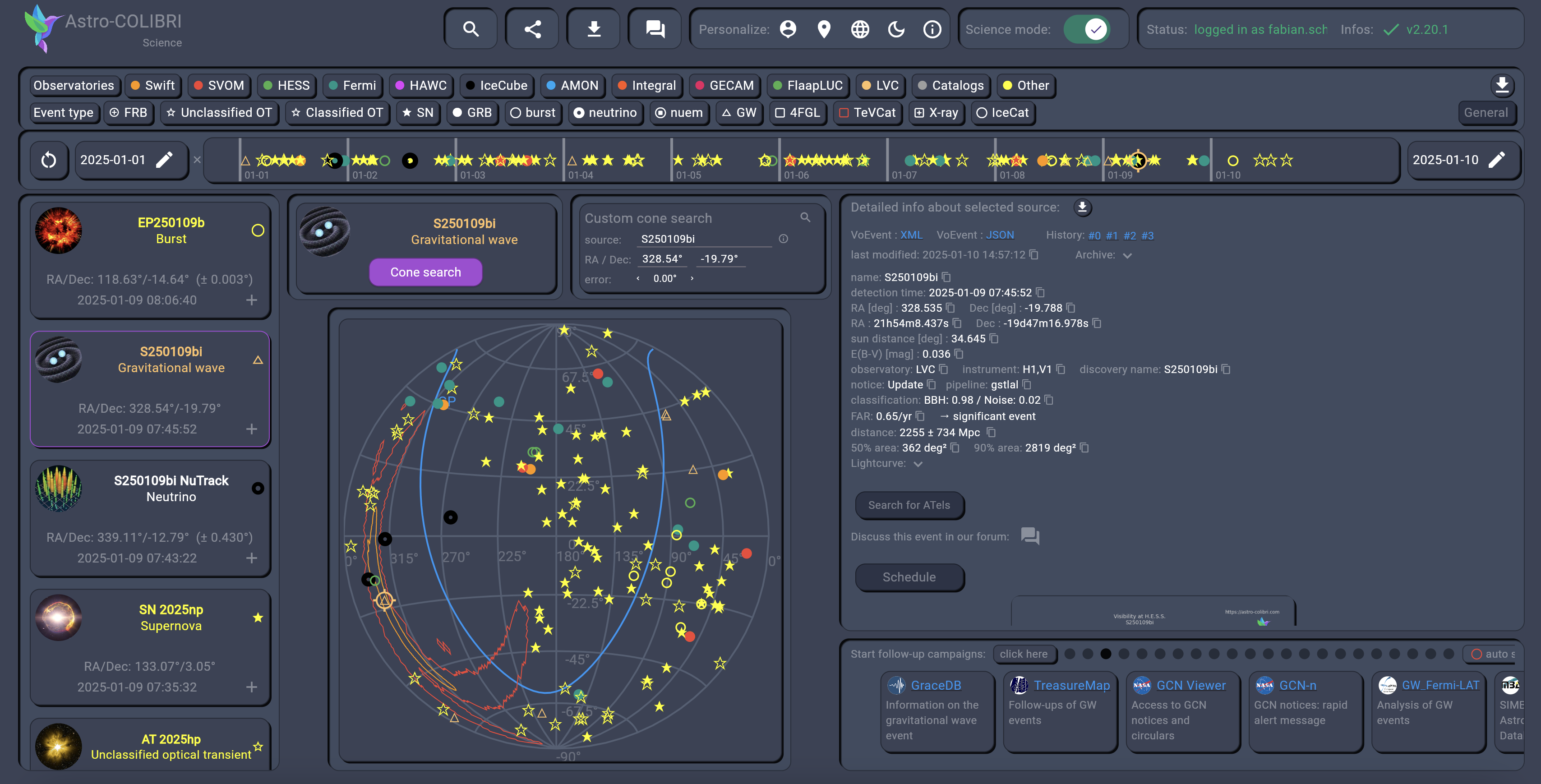}
  \caption{Web interface available at \url{https://astro-colibri.com}}
  \label{fig:sub1}
\end{subfigure}%
\begin{subfigure}{.35\textwidth}
  \centering
  \includegraphics[width=.94\linewidth]{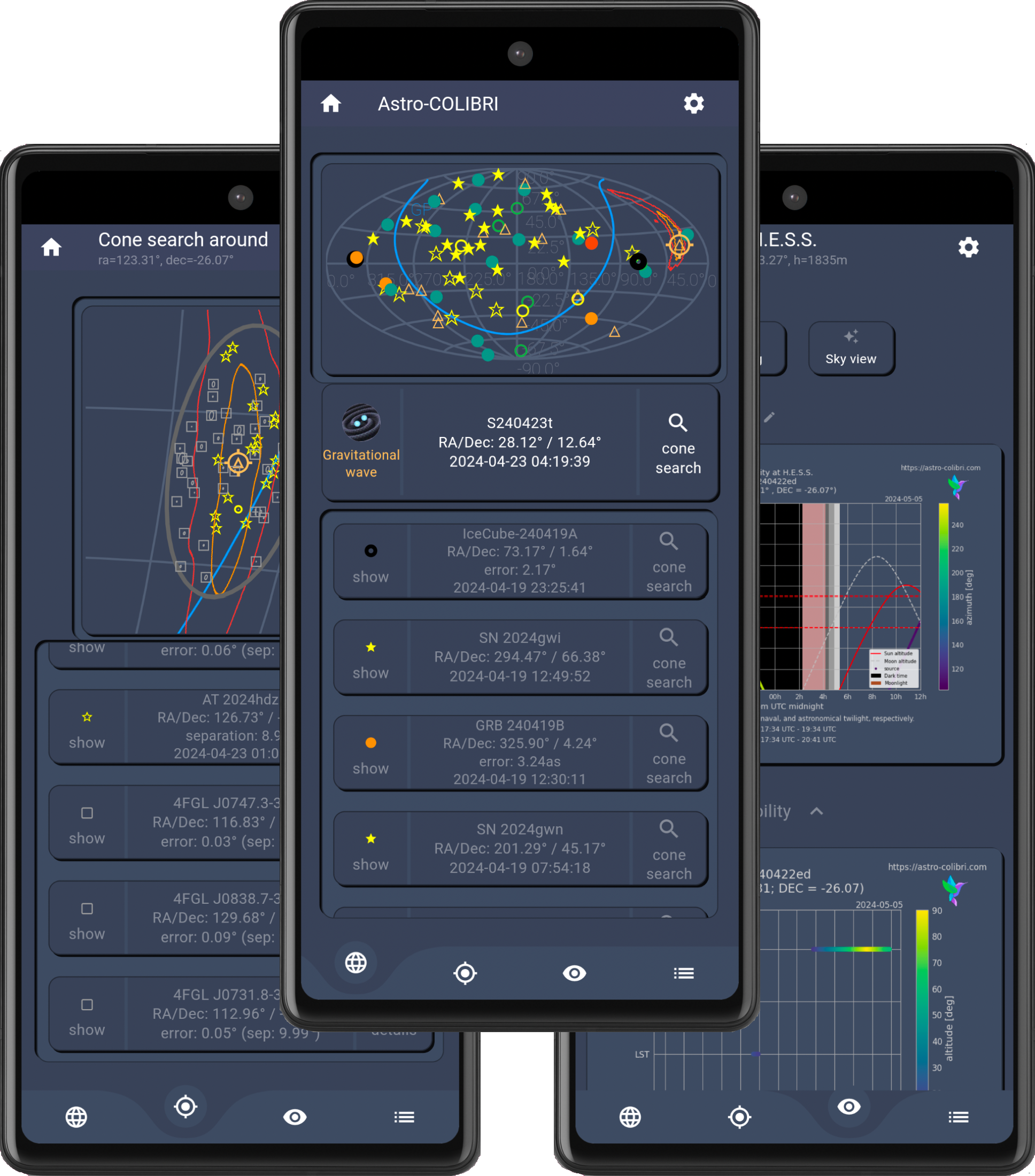}
  \caption{Android and iOS apps}
  \label{fig:sub2}
\end{subfigure}
\caption{The graphical Astro-COLIBRI user interfaces.}
\label{fig:test}
\end{figure}

\section{Lightcurve Generation}
\label{sec:workflow}

The lightcurve generation process is fully automated and designed to deliver reliable photometric data with minimal latency. The workflow consists of four main stages: data sourcing, aggregation and processing, quality filtering, and product generation.

\subsection{Data Aggregation and Processing}
\label{subsec:processing}

To build a comprehensive lightcurve, we query data from three major public surveys, each with its unique characteristics:
\begin{itemize}
    \item \textit{Asteroid Terrestrial-impact Last Alert System (ATLAS):} A robotic survey that scans the entire sky several times per night, providing prompt detection of near-Earth asteroids and other transients~\citep{Tonry2018}.
    \item \textit{All-Sky Automated Survey for Supernovae (ASAS-SN):} A network of telescopes dedicated to finding bright supernovae and other transients across the entire sky~\citep{Kochanek2017}.
    \item \textit{Zwicky Transient Facility (ZTF):} A time-domain survey of the northern sky. We access ZTF data in real-time through the \emph{Fink} broker~\citep{Moller2021}, which processes the ZTF alert stream.
\end{itemize}
Thanks to a modular setup, additional data sources can be added easily. We use this feature to provide networks of amateur astronomers a way to compare their observations with the data obtained by the above mentioned surveys.

For any given transient event, Astro-COLIBRI searches for the nearest counterpart in the catalogs of these surveys. A crucial distinction exists in the type of photometric data provided. ZTF (via Fink) and ATLAS report \textit{difference magnitudes}, which are derived by subtracting a deep sky template image from the observation. This method effectively isolates the flux of the transient from any underlying steady-state source, such as its host galaxy. In contrast, ASAS-SN provides raw (aperture) photometry. These measurements include the combined light from the transient and its host galaxy. This difference is important for correct scientific interpretation and is made clear to the user.

Once retrieved, the data undergoes minor processing. To improve clarity for high-cadence data, measurements from ATLAS are combined into 90-minute time bins, with the mean value plotted. The original data points are also shown for full transparency. To ensure the reliability of the final lightcurve, we apply several quality filters to the raw photometric data points from all observatories:
\begin{enumerate}
    \item Entries with very bright, likely spurious magnitudes (mag $< 0.1$) are removed.
    \item Measurements with large photometric uncertainties (mag\_err $> 0.5$) are excluded from further treatment.
\end{enumerate}

Data points that are flagged as poor quality by their source are handled specifically. For ZTF, measurements marked as 'bad quality' by the Fink broker are shown with semi-transparent markers. Similarly, the original (pre-binning) ATLAS measurements are displayed semi-transparently. Upper limits from all surveys are consistently represented as downward-facing triangles.

\subsection{Data Products and Availability}
\label{sec:products}

The primary outputs of this process are a graphical image of the lightcurve, a downloadable data file (CSV format), and a summary of the photometric evolution. All products are accessible through the Astro-COLIBRI user interfaces.

\begin{figure}
\centering
\begin{subfigure}{.48\textwidth}
  \centering
  \includegraphics[width=.95\linewidth]{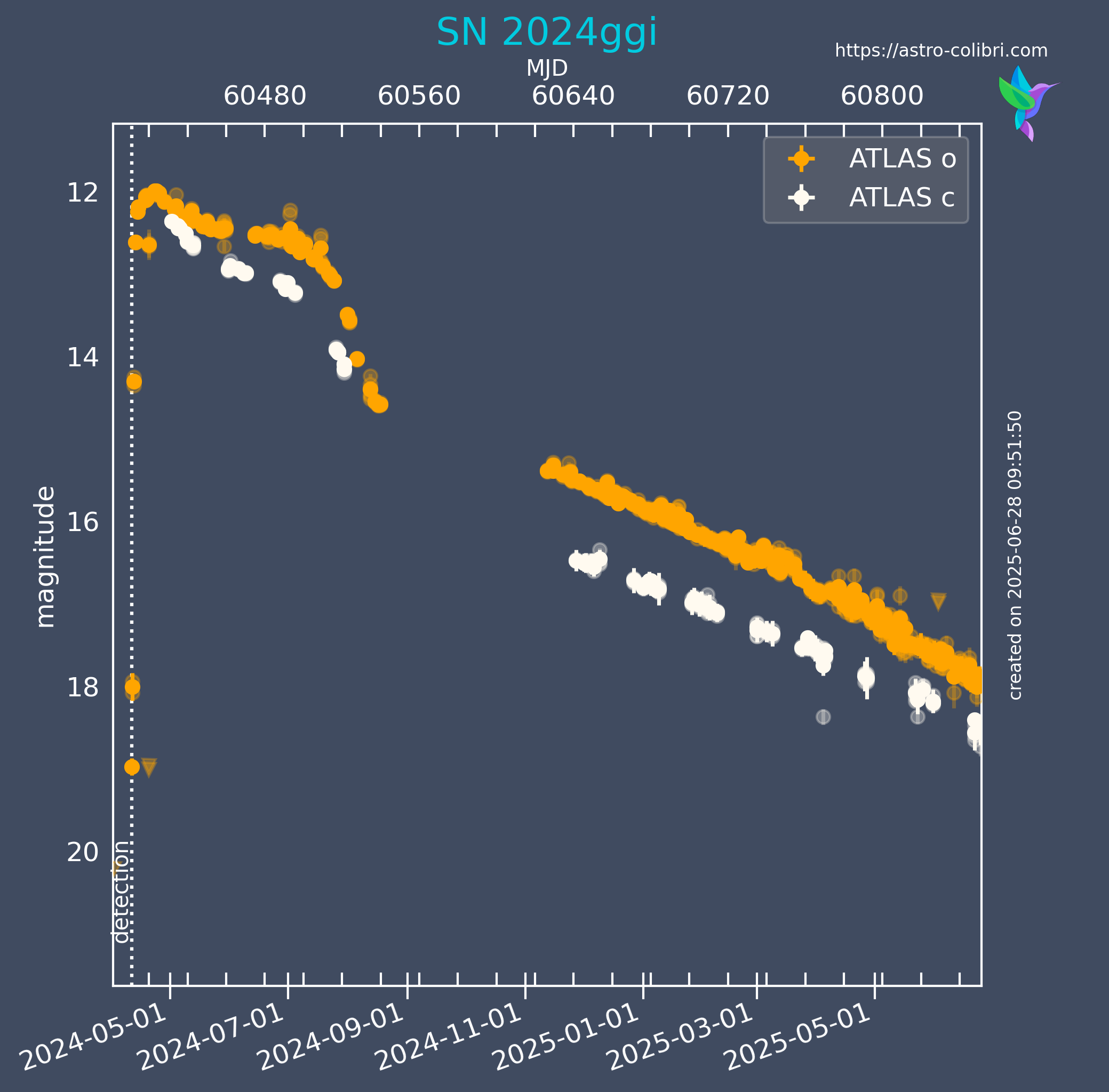}
  \caption{SN 2024ggi, a type-II supernova}
  \label{fig:SN}
\end{subfigure}%
\begin{subfigure}{.48\textwidth}
  \centering
  \includegraphics[width=.95\linewidth]{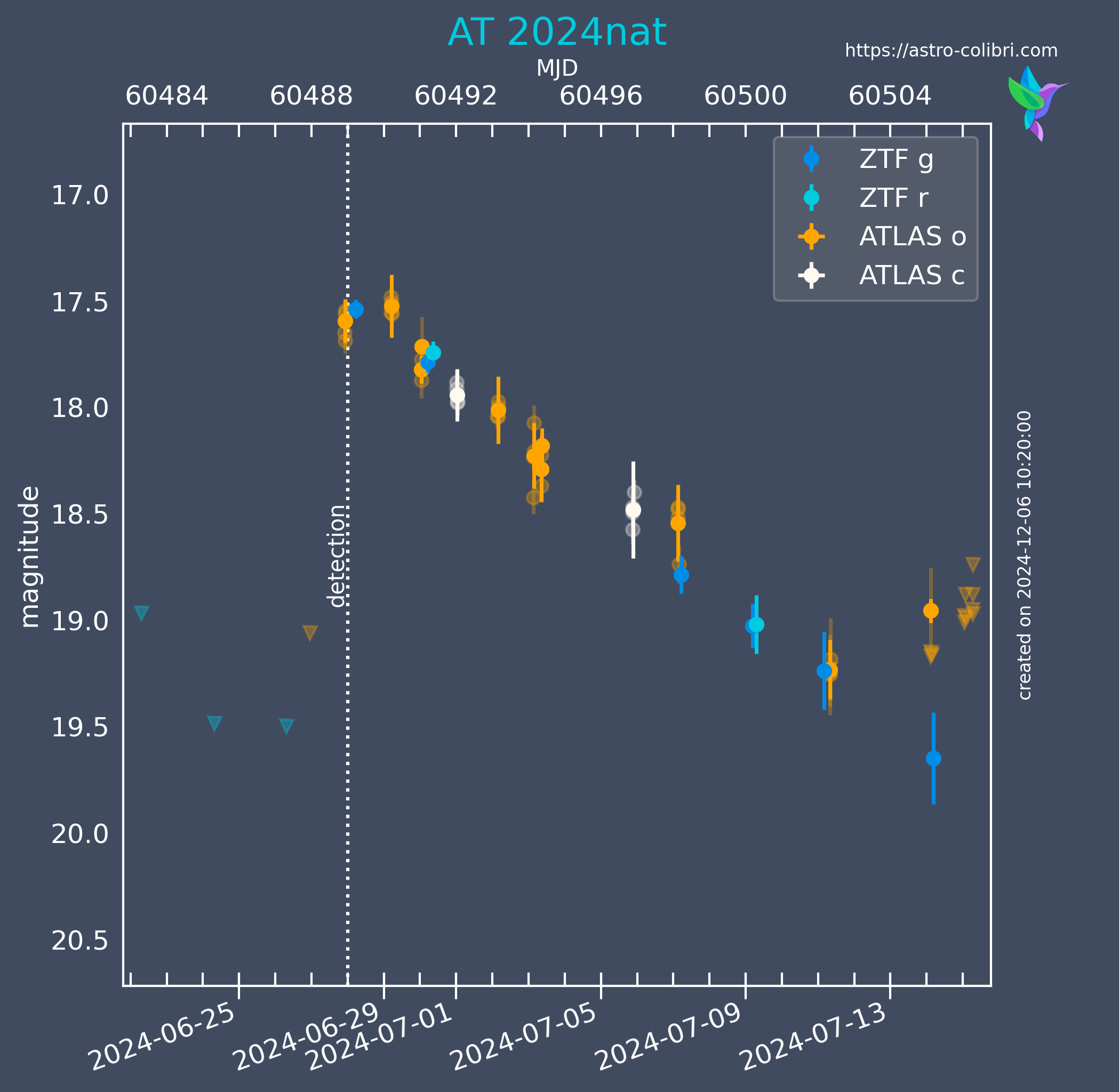}
  \caption{AT2024nat, a cataclysmic variable detected by ATLAS}
  \label{fig:ccv}
\end{subfigure}
\caption{Example of optical lightcurve generated by Astro-COLIBRI across different time ranges. Data from ATLAS (orange and white), and ZTF (blue) are shown. Semi-transparent markers indicate original ATLAS data points before rebinning.}
\label{fig:lightcurve}
\end{figure}

The final lightcurve plot provides a clear, at-a-glance summary of the transient's photometric evolution. Examples for the produced lightcurve figures are given in Fig.~\ref{fig:lightcurve}. The x-axis shows the time, automatically scaled to encompass all available data points, while the y-axis shows the magnitude. Different colors and markers are used to distinguish between the various data sources. The date of the plot's creation is watermarked on the figure, ensuring users are aware of its timeliness. 

By default, Astro-COLIBRI updates the lightcurves for all classified optical transients detected within the last 30 days twice daily. These standard plots cover a time range from 5 days before and up to 30 days after the initial detection.

Users can also generate customized lightcurves. In "Science mode", a menu in the graphical interfaces allows the selection of a user-defined time range and the choice of the survey/observatory. Users with appropriate access rights can also incorporate data from proprietary sources or amateur astronomer networks (e.g., RAPAS).

Alongside the figure, a comprehensive comma-separated value (CSV) file is available for download. This file contains all photometric measurements used to create the plot. Each row details the observation time (UTC and MJD), filter, magnitude, uncertainty, and data source. A `flag` column provides further context, indicating if a point is an upper limit (`ul`), of good or bad quality, or part of a binned average, empowering users to perform their own detailed analysis.

\subsubsection{Photometric Summary}
To provide deeper insights into the lightcurve data and to allow their use for efficient event selection and filtering, we provide key photometric data for each event. This includes three crucial data points summarizing the evolution of the lightcurve: the \textit{first} observation is defined as the earliest measurement at or after the TNS detection time (i.e., the official event report time, with a 12-hour window included to ensure better coverage). The \textit{peak} observation corresponds to the measurement of highest brightness, and the \textit{last} observation is the most recent entry in the telescope database. These values are extracted for each filter and telescope (ATLAS, ASAS-SN and ZTF) if available. This summary data serves as input to detailed filters, enabling users to identify rapidly rising or fading transients, filter events by the last magnitude value, or select events within a specific time window relative to the first, peak, or last observation.

\section{Scientific Use Case: Pinpointing Cosmic Accelerators}
The power of Astro-COLIBRI in accelerating science is best illustrated with a real-world example. In December 2021, the IceCube Neutrino Observatory detected a high-energy neutrino, IceCube-211208A~\citep{IceCube211208A}, with a high probability of being astrophysical in origin. Within hours, users of Astro-COLIBRI were able to see that the neutrino's arrival direction was spatially coincident with a gamma-ray blazar, PKS 0735+178, which was simultaneously undergoing a major flare detected by the Fermi-LAT space telescope. Additional detections of neutrinos by GVD, BUST, and KM3NeT further increased the interest of this event. 

The instant contextualization, available on the mobile phones of astronomers worldwide, allowed for the rapid triggering of an extensive multi-wavelength follow-up campaign with telescopes across the globe, including the H.E.S.S. and VERITAS gamma-ray observatories~\citep{Acharyya2023}. While this association was not conclusive, the speed at which the community was mobilized demonstrates the platform's ability to significantly accelerate the scientific process.

\section{Conclusion}
\label{sec:conclusion}
The automated generation of optical lightcurves represents a significant enhancement to the Astro-COLIBRI platform. By aggregating, processing, and visualizing data from key astronomical surveys, this feature provides an essential tool for the rapid assessment and scientific analysis of transient events. The transparent handling of data from different sources and the provision of clean plots and detailed data files empower our users to facilitate their research in the dynamic field of time-domain astronomy. The available photometric data allows for fine-grained event filtering enhancing the customization options available to the user and paving the way to an efficient use of the wealth of information that will be soon provided by the Vera Rubin Observatory / LSST.

\section*{Acknowledgements}
The authors acknowledge the support of the French Agence Nationale de la Recherche (ANR) under reference ANR-22-CE31-0012. This work was also supported by the European Union's Horizon 2020 Programme under the AHEAD2020 project (grant agreement n. 871158) and by the Horizon Europe Research and Innovation programme under the ACME project (grant agreement n. 101131928).


\end{document}